\documentclass[preprints,article,accept,pdftex,moreauthors]{Definitions/mdpi} 
\firstpage{1} 
\pubvolume{1}
\issuenum{1}
\articlenumber{0}
\pubyear{2025}
\copyrightyear{2025}
\datereceived{ } 
\daterevised{ } 
\dateaccepted{ } 
\datepublished{ } 
\hreflink{https://doi.org/} 

\DeclareMathOperator*{\argmin}{argmin}
\Title{Intrinsic motivation as constrained entropy maximization}

\TitleCitation{Title}

\Author{Alex B. Kiefer $^{1,2}$*\orcidA{}}

\AuthorNames{Alex B. Kiefer}

\isAPAStyle{%
       \AuthorCitation{Kiefer, A.B.}
         }{%
        \isChicagoStyle{%
        \AuthorCitation{Kiefer, Alex B.}
        }{
        \AuthorCitation{Kiefer, A.}
        }
}

\address{%
$^{1}$ \quad VERSES, Los Angeles, California, USA\\
$^{2}$ \quad Monash Centre for Consciousness and Contemplative Studies, Monash University, Melbourne, VIC 3800, Australia}

\corres{Correspondence: alex.kiefer@monash.edu}

\abstract{``Intrinsic motivation'' refers to the capacity for intelligent systems to be motivated endogenously, i.e. by features of agential architecture itself rather than by learned associations between action and reward. This paper views active inference, empowerment, and other formal accounts of intrinsic motivation as variations on the theme of constrained maximum entropy inference, providing a general perspective on intrinsic motivation complementary to existing frameworks. The connection between free energy and empowerment noted in previous literature is further explored, and it is argued that the maximum-occupancy approach in practice incorporates an implicit model-evidence constraint.}

\keyword{intrinsic motivation; active inference; empowerment; entropy} 

\begin{document}


\section{Introduction}

In psychology, ``intrinsic motivation'' refers to the tendency for intelligent creatures to be motivated to do certain things (such as explore, learn, and grow) even in the absence of specific external reward signals \citep{DiDomenico2017TheEN}. This paradigm has increasingly gained traction in machine learning, where it is operationalized as the idea that policies for action may be optimized based on structural features of agents and agent-environment interactions, as against traditional approaches like reinforcement learning, which optimize policies based on \emph{ad hoc} reward functions.

An early, and increasingly influential, formal account of intrinsic motivation is based on \emph{empowerment}, defined as the capacity of the channel linking agents' actions (actuator states) to sensory feedback (observations) \citep{Klyubin2005EmpowermentAU, Salge2013EmpowermentA}. One interpretation of this objective is that empowered agents ``keep their options open'', as wide action-conditioned channel capacity entails that agents are able to realize a variety of states (for which observations are a proxy).

The \emph{active inference} framework \citep{Friston2017ActiveIA} shares similar motivations, and provides a Bayesian method for combining a general form of intrinsic motivation (i.e. curiosity or ``epistemic drive'') \citep{Friston2015ActiveIA} with agent-specific prior distributions over states or outcomes \citep{dacosta2024activeinferencemodelagency}, which model homeostatic set points and can function like explicit rewards. The expected (variational) free energy (EFE), as discussed below, guides policy selection in this framework by supplying an empirical prior over policies, given observations. 

More recently, the objective of \emph{maximum path occupancy} has been proposed as a framework for intrinsic motivation \citep{RamirezRuiz2022ComplexBF}. On this account, agents are motivated to maximize future action-state path occupancy, which can be measured in terms of both the entropy of the action distribution and the entropy of the ensuing state distribution, given an initial state. This somewhat more radical perspective explicitly inverts the perhaps natural assumption that drives for exploration and curiosity have evolved as a means to achieving reward, and in effect views rewarding states as instrumentally valuable in enabling future exploration, i.e. avoiding absorbing states that afford little or no action variability (e.g. death).

There are many other formal treatments of intrinsic motivation in the literature on machine learning, some closely related to those just discussed, such as pioneering work on artificial curiosity (see e.g. \cite{Schmidhuber1991AdaptiveCA, Schmidhuber2010FormalTO}) and treatments in terms of Bayesian surprise \citep{ITTI20091295, Mazzaglia2021CuriosityDrivenEV}. Here, the focus is mainly on the relationship between active inference and empowerment, and on the relationship of both to maximum occupancy which has recently been proposed explicitly as an alternative.

While \citep{moreno-bote2023empowerment} conducts a comparative empirical study of these three frameworks for intrinsic motivation on a toy problem and \citep{Biehl2018ExpandingTA} considers how active inference may be formally related to broader schemes for intrinsic motivation, comparatively little work exists on the formal and conceptual relations among these frameworks. Here, I highlight the fact that all three can be understood as variations on the theme of constrained entropy maximization, a principle with deep connections to the free energy principle and active inference  \citep{sakthivadivel2022geometryanalysisbayesianmechanics}. I explore the connection between empowerment and active inference \citep{10.1162/neco_a_01351} by casting the empowerment objective itself explicitly as a form of variational inference. I also argue that the ability of occupancy-maximizing agents to exhibit apparently goal-directed behavior depends on a ``survival instinct'' or model-evidence constraint implicit in the factorization of the overall system into actions and states. These considerations frame entropy maximization, under local constraints, as the kernel of intelligence and agency, with particular facets of this process such as empowerment, perception, curiosity and the ``will to live'' as corollaries.

The first section below unpacks the three frameworks for intrinsic motivation mentioned above (empowerment, active inference, and maximum occupancy) in some detail, both formally and in terms of conceptual motivation, and articulates their ties to constrained entropy maximization. Section two looks closely at some connections among these theories, then distills a few general conclusions.

\section{Three formal accounts of intrinsic motivation}

\subsection{Empowerment}

The \emph{empowerment} objective for intrinsic motivation, originally proposed in \citep{Klyubin2005EmpowermentAU}, is defined as the capacity of the information channel linking an agent's actions to its observation of the effects of those actions. That is, given a space of possible observations $O_{T}$ at future timestep $T$ and a sequence of actions $A_{t:T}$ from the present timestep $t$ to the future, there is some distribution $P(O_{T}|A_{t:T})$ capturing the probabilistic dependence of the future observation on the actions taken, and the empowerment $\mathcal{E}$ of an agent is measured as the capacity $C$ of the information transmission channel defined by this distribution:

$$\mathcal{E}_t = C\Big(P(O_T|{A_{t:T}})\Big) = \max_{P(A)} I({A_{t:T}}; O_T)$$
In this case, $C$ is defined as the maximum mutual information between actions and future observations, $I(A_{t:T; O_T})$, when the conditional distribution $P(O_{T}|A_{t:T})$ is held fixed and the distribution over actions, $P(A)$, is allowed to vary.

It is worth taking a moment to unpack this, as a detailed understanding will be useful for comparisons below. The mutual information is standardly defined, for two random variables $X$ and $Y$, as a double sum equivalent to the KL divergence from the joint density $P(X, Y)$ to the product of the marginals over $X$ and $Y$:

\begin{linenomath}
\begin{equation*}
\begin{split}
I & = \sum_{y \in \mathcal{Y}} \sum_{x \in \mathcal{X}} P(x,y) \log \Big( \frac{P(x,y)}{P(x)P(y)}\Big)\\
& = D_{KL}\Big(P(X,Y)||P(X)P(Y)\Big)
\end{split}
\end{equation*}
\end{linenomath}

Intuitively, this expression measures how different the actual joint distribution is from what it would be were the two variables independent, i.e. how much information the variables carry about one another. While this measure is symmetric (i.e. the same for $X$ and $Y$), it can be broken down in terms of conditional probabilities in either direction. Since the joint density can be factorized into a prior and a conditional density, i.e. $P(X,Y) = P(X)P(Y|X) = P(Y)P(X|Y)$, the mutual information can also be expressed as an expected KL divergence from a conditional density $P(Y|X)$ to the marginal over $Y$:

\begin{linenomath}
\begin{align*}
I & = \sum_{y \in \mathcal{Y}} \sum_{x \in \mathcal{X}} P(x)P(y|x) \log \Big( \frac{P(x)P(y|x)}{P(x)P(y)}\Big) && \text{Factorize joint distribution}\\
& = \sum_{x \in \mathcal{X}} P(x) \Bigg[\sum_{y \in \mathcal{Y}} P(y|x) \log \Big( \frac{P(y|x)}{P(y)}\Big) \Bigg] && \text{Cancel out $P(x)$s and rearrange} \\
& = \mathbb{E}_{P(X)} \Bigg[D_{KL}\Big(P(Y|X)||P(Y)\Big) \Bigg] &&
\end{align*}
Given a fixed $P(Y|X)$ (channel), the \emph{channel capacity} $C\Big[P(Y|X)\Big]$ is then the maximum value this mutual information can take, given a free choice of $P(X)$.
\end{linenomath}

The empowerment objective is just this channel capacity, with respect to the channel linking actions over timesteps $t \dots T$ with observations at $T$.\footnote{For simplicity the discussion here focuses on the original formulation in \citep{Klyubin2005EmpowermentAU}, but obviously many variations on this theme are possible, e.g. using different time indices (as explored in \citep{PRXLife.2.033009}) or swapping out observations for latent states.} Intuitively, the mutual information term (i.e. information gain expected under the action distribution) measures both the \emph{controllability} of outcomes (the influence of action selection on such outcomes) and the \emph{variety} of achievable outcomes (i.e. ``keeping one's options open'') \citep{Klyubin2005EmpowermentAU}. This combination of controllability and variety is characteristic of constrained entropy maximization, a common theme in many frameworks for intrinsic motivation \citep{PRXLife.2.033009, PhysRevLett.110.168702, RamirezRuiz2022ComplexBF}, and is related to Ashby's ``law of requisite variety'' \citep{Ashby1991}.

The ``variety'' aspect of empowerment can be made more explicit by considering the relation of mutual information to entropy. Any mutual information $I(X;Y)$ can be expressed in terms of entropy in several ways:

\begin{linenomath}
\begin{align*}
I(X;Y) & = H(X) - H(X|Y) \\
& = H(Y) - H(Y|X) \\
& = H(X) + H(Y) - H(X,Y)
\end{align*}
\end{linenomath}
Thus empowerment can be seen as maximizing the entropy of the action distribution $H\big(P(A)\big)$ while ensuring that actions are ``rational'' in the sense of being reliably related to observations, i.e. minimizing $H\big(P(A|O)\big)$. At the same time, it can be viewed as maximizing the variety of observations $O$ while ensuring that they remain controllable, i.e. minimizing $H\big(P(O|A)\big)$.\footnote{There is a \emph{prima facie} conflict here with the reasonable objective of maximizing model evidence (i.e. minimizing the surprisal of observations), as in active inference. This matter is discussed further below, but note that in active inference treatments the model evidence (marginal likelihood) is usually treated as fixed at the timescale of inference, and agents simply act so as to furnish evidence for this model.} 

This objective may be read as a signal guiding model \emph{evolution} or selection, as in the work just cited (i.e. choosing a generative model of actions and outcomes $P(A)P(O|A)$). Given a fixed model, agents may also choose policies (actions) so as to maximize the time-dependent empowerment $\mathcal{E}_t$ by seeking the position in the state-space of the overall system (where external states are implicitly represented here by observations) in which the channel capacity is highest, since $P(O_T|{A_{t:T}})$ implicitly depends on the states at $t...T$.

Before moving on to consider other treatments of intrinsic motivation, we note that in \citep{PRXLife.2.033009} it is shown (in the setting of continuous state-spaces) that generalizing the empowerment objective just discussed, by varying the length of action and observation sequences and the time interval separating actions from target observations, allows one to recover various extant descriptions of control in dynamical systems. Saliently for present purposes, a generalized empowerment objective in which actions are taken only at the first time-step corresponds to a ``kicked'' (controlled) version of Causal Entropic Forcing \citep{PhysRevLett.110.168702}, a more general framework that models intelligent behavior in terms of entropy maximization.

\subsection{Active inference and expected free energy}

Among the most promising approaches to intrinsic motivation are those that leverage Bayesian 

Advances in cognitive (neuro)science over the past decade or so have seen the rise to prominence of the idea that most (if not all) intelligent action can be understood in terms of Bayesian inference \citep{Hohwy2013-HOHTPM-2}. This paradigm encompasses quite specific models of neuronal information processing such as predictive coding \citep{Rao1999PredictiveCI, salvatori2024a}, which has been invoked to explain perceptual inference \citep{alma99398373502466}, as well as more abstract and general frameworks, mosty saliently the free energy principle \citep{Friston2019AFE, FRISTON202335}, an account of self-organization in terms of variational Bayesian inference, and active inference \citep{Friston2017ActiveIA, SMITH2022102632}, which derives a scheme for action (i.e. planning or policy selection) from the assumption that agents select actions that are expected to minimize variational free energy in the future.\footnote{Accounts of motor control in terms of high-precision kinesthetic predictions \citep{Brown2011ActiveIA} are closely related to active inference, but here the latter term is reserved to denote the idea that policies for action are selected on the basis of expected (variational) free energy.} 

Agents governed by active inference implement a specific form of planning as inference \citep{BOTVINICK2012485}, ``reasoning backward'' from preferred outcomes (cast in this context as observations that furnish evidence for a prior generative model \citep{Hohwy2014-HOHTSB-2}) to the policies most likely to bring them about. In brief, this involves inferring a (variational) posterior distribution $Q(\pi)$ over policies $\pi$ in which the probability assigned to each policy is proportional to its associated model evidence. Actions are then sampled at each timestep based on a Bayesian model average of the policies, each of which entails distinct action-conditioned state transition probabilities.

The core quantity driving policy selection in the active inference framework is the expected free energy (EFE, denoted $\mathbf{G}$ in equations), which is the cumulative variational free energy that the agent expects to be incurred by choosing a policy (action sequence), given its generative model. The generative model that figures in policy inference includes a state-independent distribution $P(o)$ over outcomes (observations $o$) that the agent ``prefers'' to see, which can be cast as the marginal likelihood of observations \citep{dacosta2024activeinferencemodelagency} and models the characteristic attracting set of states that homeostatic systems must remain within in order to persist \citep{Friston2019AFE}. This can be thought of as a kind of intrinsic motivation, since it is ``built in'' to the agent rather than learned, though in practice (i.e. in computational models) it functions similarly to an \emph{ad hoc} reward function. Crucially, however, the EFE also implements the model-independent inductive bias that actions will minimize variational free energy in the future, and thus subserves a more general form of intrinsic motivation.

The EFE associated with a policy $\mathbf{G}_\pi$ is defined as the expectation of the variational free energy, given the state-transition probabilities induced by following that policy \citep{Friston2015ActiveIA}. This depends on possible future observations, which are assumed to be generated independently by states at each timestep, so that $\mathbf{G}_\pi$ can be computed as a sum over timestep-specific terms $\mathbf{G}^t_\pi$. Selection of actions or control states $u$ can then be summarized as follows, where $P(u_t = u_i|\pi_j)$ is 1 if policy $j$ begins with control state $u_i$ and 0 otherwise, and $\mathbf{F}_\pi$ is the variational free energy (VFE) incurred by policy $\pi$:\footnote{I omit several features of active inference models that are inessential for present purposes, such as the baseline policy or ``habit'' prior and the temperature parameter used in action selection. Please see \citep{Friston2017ActiveIA,Heins2022, SMITH2022102632} for further details. In some treatments, $\mathbf{F}$ is omitted as well.}

\begin{linenomath}
\begin{align*}
 u_t \sim Q(u_t) & = \sum_\pi P(u_t|\pi)Q(\pi) \text{\hspace{0.47cm} Control state sampled from marginal}\\
Q(\pi) & = \sigma \Big(-\mathbf{G} - \mathbf{F}\Big) \text{\hspace{0.94cm} Posterior over policies} \\
\mathbf{F}_\pi & = \sum_{t=0}^T \underbrace{\mathbb{E}_{Q(s_t|\pi)} \big[ - \log P(s_t, o_t | \pi)\big]}_{\text{Energy}} - \underbrace{H\big[Q(s_t|\pi)\big]}_{\text{Entropy}} \text{\hspace{0.4cm} VFE of policy $\pi$} \\
\mathbf{G}_\pi & = \sum_{t=0}^T \mathbf{G}^t_\pi \text{\hspace{6.15cm} EFE of policy $\pi$} \\
\end{align*}
\begin{align*}
    \mathbf{G}^t_\pi & = \mathbb{E}_{Q(s_t, o_t|\pi)} \Big[ \log Q(s_t|\pi) - \log P(s_t, o_t|\pi)\Big] && \\
& \approx \mathbb{E}_{Q(s_t, o_t|\pi)} \Big[ \log Q(s_t|\pi) - \log Q(s_t|o_t, \pi) - \log P(o_t)\Big] && \\
& = -\underbrace{\mathbb{E}_{Q(s_t, o_t|\pi)} \Big[ - \log P(o_t)\Big]}_{\text{Expected utility}} - \underbrace{D_{KL}\Big[Q(s_t|o_t, \pi) || Q(s_t|\pi)\Big]}_{\text{Information gain}} &&
\end{align*}
\end{linenomath}

Given that the EFE is defined as the VFE expected under various policies, it seems at first glance that considering both $\mathbf{F}$ and $\mathbf{G}$ when computing $Q(\pi)$ involves double-counting. The crucial difference is that the EFE is used to compute a \emph{belief} $P(\pi) = \sigma\big(-\mathbf{G}\big)$ about policies, which is used as a prior in the full variational policy inference scheme \citep{Heins2022}:

\begin{linenomath}
\begin{align*}
Q^*(\pi) & = \argmin_{Q(\pi)} \mathcal{F} \\
\mathcal{F} & = \Bigg[ D_{KL}\Big(P(\pi)||Q(\pi)\Big) + \mathbb{E}_{Q(\pi)}\Big[\mathbf{F}_\pi\Big]\Bigg]
\end{align*}
\end{linenomath}
where $\mathcal{F}$ is the total variational free energy and $Q^*(\pi)$ is the optimal variational posterior over policies.

The role of the EFE as a prior is underwritten by two differences with respect to the VFE. First, as discussed in \citep{10.1162/neco_a_01354}, the variational distribution over states in the EFE, $Q(s_t|\pi)$, is a variational \emph{empirical prior}---computed by conditioning on the most recently inferred state distribution $Q(s_t)$ and rolling the generative model out into the future---as opposed to the variational posterior $Q(s_t)$ over states, which inverts the likelihood while incorporating the prior over states. The appearance of $Q(s_t|\pi)$ in the EFE (together with the use of $Q(s_t|o_t, \pi)$ to approximate $P(s_t|o_t)$) underwrites the EFE's information gain term.

A second difference is that the marginal over observations $P(o_t)$, rather than the likelihood $P(o_t|s_t)$, appears in the EFE, in accord with the EFE's role as a prior belief about which policies should be (i.e. will be, in a planning-as-inference scheme) pursued.\footnote{Importantly, while some formulations (e.g. \citep{10.1162/neco_a_01354}) as well as many simulations employ a ``preference distribution'' $\tilde{P}(o_t)$ specified independently of the predictive (generative) model of the world, this is not a deep feature of active inference. In \citep{dacosta2024activeinferencemodelagency} for example the EFE objective is formulated solely in terms of the difference between $P(o,s)$ and $P(o,s|a)$, i.e. the ``reward`` or preference model is the same generative model used for prediction, with actions marginalized out.} The per-policy variational free energy $\mathbf{F}$, on the other hand, takes into consideration the entropy of the posterior state distribution as well as the expected energy of observations under the likelihood, in such a way that the entropy of the posterior is maximized under the energy (model evidence) constraint, in accordance with the principle of constrained maximum-entropy inference \citep{  PhysRev.106.620, Friston2019AFE, Kiefer2020-KIEPIA, sakthivadivel2022geometryanalysisbayesianmechanics}.

\subsection{Maximum occupancy}

The Maximum Occupancy Principle (MOP) \citep{RamirezRuiz2022ComplexBF} carries the theme of intrinsic motivation to its logical conclusion, proposing that a traditional picture of rational agency, in which curiosity and other intrinsic drives have evolved in order to serve reward maximization, should be inverted: we can instead understand rewarding states as a means to the end of continuing to live, i.e. to explore (thus maximally occupy) action-state path space.

Formally, the occupancy objective is defined in terms of a state-conditioned policy distribution $\pi(A|S)$ and transition dynamics $P(S'|S, A)$, which can be alternately sampled from to generate action-state paths $\tau$. The reward function $R(\tau)$ for a given trajectory is then specified as:

\begin{linenomath}
\begin{align*}
R(\tau) & = - \sum_{t=0}^\infty \gamma^t\log \Big[ \pi^{\alpha}(a_t|s_t)P^{\beta}(s_{t+1}|s_t, a_t)\Big]
\end{align*}
\end{linenomath}
where $\gamma^t$ is the standard temporal reward discount in reinforcement learning and $\alpha$ and $\beta$ are weights modulating the influence of action and state path occupancy. Agents select policies so as to maximize the expected reward or ``value'' of states $s$, $V_{\pi}(s)$:

\begin{linenomath}
\begin{align*}
V_{\pi}(s) & = \mathbb{E}_{\pi(A|S)P(S'|S,A)} \Big[R(\tau) | s_0 = s\Big] \\
& = \mathbb{E}_{\pi(A|S)P(S'|S,A)} \Bigg[\sum_{t=0}^\infty \gamma^t \Big( \alpha H(A|s_t) + \beta H(S'|s_t, a_t) \Big) | s_0 = s \Bigg]
\end{align*}
\end{linenomath}

Here, the realization of $\tau$ depends on the initial state $s$, and $H(A| s_t)$ and $H(S'|s_t, a_t)$ denote the conditional entropy of the action distribution given the current state, and of the distribution of the next state given the current state and action, respectively.\footnote{Note that, while the expectation over conditioned variables is absorbed in the entropy terms in the third line, the expectation over the conditioning variables $s_t$ and $a_t$ remains.} Thus, agents that maximize $V_{\pi}(s)$ maximize an expectation over the (summed step-wise conditional) entropy of both action and state paths, subject to the weights and initial condition.\footnote{\emph{Negative} weights can give rise to entropy-minimizing behavior as well, as discussed below.}

In \citep{moreno-bote2023empowerment}, empirical studies are presented in which MOP agents aggressively explore state and action space while still exhibiting apparently goal-directed behavior. The former is perhaps to be expected, given the purely intrinsic, surprisal-maximizing reward function, thanks to which agents will directly seek out improbable actions that lead to improbable states. Presumably, the ability of MOP agents to behave in goal-oriented ways despite the absence of explicit tasks, rewards, or even preference distributions, is underwritten by the imperative to maximize longer-term path occupancy,\footnote{Notably, similar emergent task-oriented behavior is demonstrated in agents governed by an empowerment objective in \citep{ringstrom2023rewardnecessarycreatemodular}.} which balances the tendency to greedily maximize entropy at each timestep. This implicit constraint on short-term entropy maximization in the service of increasing entropy in the long run is evocative of the argument in \citep{Ueltzhffer2020OnTT} according to which the structured, relatively low-entropy states characteristic of complex forms of life are favored for their ability to accelerate the dissipation of free energy within the broader universe.

\section{A unified view of intrinsic motivation}

This section begins by analyzing the relationship between active inference and empowerment, then considers the maximum-occupancy perspective in relation to both of these. It then concludes with a discussion of some themes common across these frameworks, and a synthesis that allows us to resolve some apparent dichotomies from a multi-scale or scale-free perspective.

\subsection{Empowerment and active inference}

Maximizing the empowerment objective is closely related to minimizing expected free energy. Most straightforwardly, in the absence of a constraint (expected utility term), the expected free energy described above reduces to the negative information gain $D_{KL}\Big(Q(s_t|o_t, \pi)||Q(s_t|\pi)\Big)$, so that minimizing EFE maximizes the mutual information between states and observations \citep{FRISTON2024108891}.\footnote{Again, even though this mutual information can be interpreted as maximizing the entropy of observations (constrained by their controllability), optimization of $Q$ in variational inference is always constrained by the generative model $P$ so there is no conflict with the imperative to maximize model evidence. That said, in a multi-scale setting one may also consider learning the parameters of $P$ as discussed below.}

While the original empowerment objective \citep{Klyubin2005EmpowermentAU} leaves the mediation of the action-sensation channel $P(O_T|A_{t:T})$ by hidden states implicit, the active inference objective simply makes this explicit: in choosing actions, agents effectively choose the transition dynamics for controllable states (in typical implementations, discrete actions index slices of transition tensors), such that they are rendered informative about observations. Thus effectively, states are (probabilistically) chosen so as to maximize the mutual information between actions and observations, as in the empowerment objective. 

In \citep{10.1162/neco_a_01351} (Appendix), it is claimed that ``empowerment is a special case
of active inference, when we can ignore risk (i.e., when all policies are equally risky)''. Here, \emph{risk} is a term occurring in the following alternative breakdown of the EFE (see \citep{SMITH2022102632}, Appendix A for a derivation):

$$\mathbf{G}^t_\pi = \underbrace{D_{KL}\Big( Q(o_t|\pi)||P(o_t)\Big)}_{\text{Risk}} + \underbrace{\mathbb{E}_{Q(s_t|\pi)}\Big[H(P(o_t|s_t))\Big]}_{\text{Ambiguity}}$$

Intuitively, risk is simply a measure of expected negative reward, which in this context is how different predicted outcomes are from those expected \emph{a priori} (i.e. preferred). The entropy of the likelihood mapping from states to observations expected under a given policy (``Ambiguity'') quantifies how uncertain the agent will be about outcomes if that policy is pursued. Thus, minimizing expected free energy encourages agents to choose policies (actions) that render outcomes predictable, subject to the constraint that risk is minimized.

We can run a similar argument by considering the empowerment objective described in \citep{Klyubin2005EmpowermentAU} as part of a variational inference process. In terms of the notation used for active inference, the goal would be to maximize $I_t(\pi; o_T)$, where as above $\pi$ is a sequence of control states $[u_0, u_1,\dots, u_T]$. This objective can be expressed in terms of the entropies of posterior observation and policy distributions, and also as a KL divergence:

\begin{linenomath}
\begin{align*}
I_t(\pi; o_T) & = H\Big(Q(\pi)\Big) - H\Big(Q(o_T|\pi)\Big)\\
& = D_{KL}\Big(Q(\pi, o_T)||Q(\pi)Q(o_T)\Big) \\
\end{align*}
\end{linenomath}

The divergence simply states that agents maximizing empowerment should select policies that provide information about the target observation, which in this context amounts to the former affording control over the latter. The subscript in $I_t$ indicates that, like the original empowerment objective $\mathcal{E}_t$, this term is implicitly time-dependent. More specifically, in the present setting, the variational posteriors $Q$ at $t$ depend on the observation $o_t$ via the state posterior $Q(s_t)$.

Interestingly, defining a conditional ``energy'' term as the negative log probability of the observation at $T$ given policies, the expression of the mutual information in terms of entropies can be written in a form analogous to a free energy $F_t(\pi,o_T)$ simply by flipping the sign and rearranging terms:
\begin{linenomath}
\begin{align*}
F_t(\pi, o_T) & = - \Big[H\Big(Q(\pi)\Big) - H\Big(Q(o_T|\pi)\Big)\Big] \\
& = \underbrace{\mathbb{E}_{Q(o_T|\pi)}\Big[- \log Q(o_T|\pi)\Big]}_{\text{``Energy''}} - \underbrace{H\Big(Q(\pi)\Big)}_{\text{Entropy}} \\
& = \underbrace{\mathbb{E}_{Q(o_T|\pi)}\Big[- \log P(o_T|s_T) \Big]}_{\text{Energy of final observation}} + \underbrace{H\Big(Q(s_T|\pi)\Big)}_{\text{Conditional state entropy}} - \underbrace{H\Big(Q(\pi)\Big)}_{\text{Policy entropy}} \\
& = \underbrace{\mathbb{E}_{Q(s_T|\pi)}\Big[H\Big(P(o_T|s_T)\Big) \Big]}_{\text{Ambiguity}} 
- \underbrace{I_t(\pi; s_T)}_{\text{State information gain}} \\
Q(s_T|\pi) & = \sum_{s_0 \in S} \dots \sum_{s_{T-1} \in S}\Big[\prod^{T-1}_{t=0}P(s_{t+1}|s_t, \pi)Q(s_0)\Big]
\end{align*}
\end{linenomath}

Maximizing $I_t(\pi; o_T)$ is then equivalent to minimizing this energy. The second line lacks the form of a proper (variational) free energy because the ``energy'' term is just the entropy of a variational density $Q(o_T|\pi)$, rather than a joint probability (generative model) $P(o, s)$. However, $Q(o_T|\pi)$ factors into several terms some of which are distributions of the generative model. Taking this into account, we arrive at the expression in the penultimate row, which is similar to a Helmholtz free energy with an additional entropy term to be \emph{minimized}: under this objective, agents will seek low-energy (predictable) observations, while maximizing the entropy of policies (``keeping options open'') and also seeking policies that minimize the entropy of the final state, i.e. seeking paths that result in controllable states. 

Finally (last line), the expected energy (negative log probability under the generative model) of $o_T$ is equivalent to the ambiguity term in the EFE mentioned above (with respect to the final observation in a trajectory), while the two entropy terms can be combined into a state information gain term.\footnote{A similar formulation of empowerment in terms of free energy is reached by considering action-state empowerment in the context of the generalized free energy functional \citep{Parr2018GeneralisedFE}, in \citep{10.1162/neco_a_01351}.} Thus from the empowerment objective alone (and ignoring additional ``preference'' constraints) we can derive drives for both \emph{epistemic value} (minimizing ambiguity) and \emph{control} (maximizing state infogain).


Active inference agents are thus ``empowered'' in that they maximize the entropy of future state distributions, under the constraint that these states or the ensuing observations be controllable. Crucially, in active inference, agents are also constrained to maximize \emph{model evidence} (or its tractable lower bound, variational free energy) \citep{Hohwy2014-HOHTSB-2}. In fact, the latter (approximately maximizing model evidence) is the central concept in the FEP and active inference, where (constrained) entropy maximization falls out of variational free energy minimization, and specifically exploratory behavior emerges thanks to the distribution-matching (KL-divergence) term in the EFE objective \citep{Millidge2021UnderstandingTO}. 

\subsection{Constrained maximum occupancy}

\emph{Prima facie}, it is difficult to square the maximum occupancy objective with those just considered in precise terms, since its objective involves only maximizing (expected) entropy, without constraints. In fact, the MOP objective described above is general enough to encode an approximation to empowerment, if the $\beta$ term is set to a negative value \citep{RamirezRuiz2022ComplexBF}, which encourages agents to choose actions that \emph{minimize} the entropy of the state transition distribution, while still maximizing the entropy of actions. This is clearly closely related to the empowerment objectives discussed above once the distinction between states and observations is accommodated (i.e., it results in agents that ``keep options open'' while ensuring controllable states and thus observations). However, while of practical interest, this really amounts to a departure from the spirit of MOP.

It is argued in \citep{RamirezRuiz2022ComplexBF} on both conceptual and experimental grounds that MOP agents exhibit more robust exploratory behavior, and variety in policy selection, than agents governed by empowerment or EFE objectives. The experiments reported in that work and in \citep{moreno-bote2023empowerment}, however, involve full observation of the state-space, so that the ambiguity term in the EFE does no work (and more generally, the usual motivations for the FEP and active inference, in which agents are assumed to infer unknown states of the environment, do not apply). Moreover, the experiments reported in \citep{RamirezRuiz2022ComplexBF} use a setting of $\beta = 0$ by default, thus effectively maximizing the entropy of only the action distribution. For these reasons the ensuing discussion focuses on the conceptual arguments surrounding entropy maximization and the role of constraints, rather than on these experimental results. 

On conceptual grounds, the MOP objective may (it is argued in \citep{RamirezRuiz2022ComplexBF}) be expected to produce a greater variety of actions than active inference for two reasons: (a) the EFE objective contains an explicit ``preference'' term which MOP lacks, and which biases action in favor of certain outcomes (thus reducing the entropy of action-state paths); and (b) while the EFE objective maximizes the entropy of the state-transition distribution at each timestep\footnote{Minimizing the variational free energy $\mathbf{F}$ in the full variational policy inference scheme maximizes the entropy of the state distribution at each step under an evidence constraint. The EFE at time $t$ can be written as a Helmholtz free energy, showing that the same is true for the policy-conditioned empirical prior over states $Q(s_t|\pi)$: \begin{linenomath}
\begin{align*}
    \mathbf{G}^t_\pi & = \underbrace{\mathbb{E}_{Q(s_t, o_t|\pi)} \Big[- \log P(s_t, o_t|\pi)\Big]}_\text{Energy} - \underbrace{H\Big(Q(s_t|\pi)\Big)}_{\text{Entropy}} && \\
\end{align*}
\end{linenomath}}, it contains no term to maximize the entropy of the action distribution.

The maximization of action (policy) entropy does seem to fall out of the empowerment framework. Thus, given the equivalences outlined above, the same should be true of active inference. \citep{RamirezRuiz2022ComplexBF} argue that the EFE deterministically selects a single policy. However, in the context of a full variational inference treatment (i.e. planning-as-inference), the entropy of the \emph{policy} distribution should also be maximized (under relevant constraints). 

Conceptually, $\pi$ is a latent variable, and \emph{ceteris paribus} its entropy should be maximized during variational inference just as the entropy of $Q(s)$, the variational density over hidden causes, is maximized. This is captured formally in work on active inference exploring a formulation of expected variational free energy that is in some ways more parsimonious than the EFE, called the \emph{generalized free energy} \citep{Parr2018GeneralisedFE}. As shown in \citep{10.1162/neco_a_01351}, this objective can (as is usual in variational inference) be written as a Helmholtz free energy, where in this case the energy term is the expected EFE under the policy posterior, and the entropy of the policy distribution is explicitly maximized as free energy is minimized:
\begin{linenomath}
\begin{align*}
    \underbrace{F\Big[Q(s,\pi)\Big]}_{\text{Generalized free energy}} &= \underbrace{\mathbb{E}_{Q(\pi)}\Big[\mathbf{G}_{\pi}\Big]}_{\text{Expected EFE}} - \underbrace{H\Big(Q(\pi)\Big)}_{\text{Policy entropy}}
\end{align*}
\end{linenomath}

It is the generalized free energy which in \citep{10.1162/neco_a_01351} is shown to be equivalent to empowerment constrained by risk. Relatedly, the ``free energy of empowerment'' $F_t(\pi, o_T)$ defined above also contains this policy entropy term. Thus, while a focus exclusively on the EFE is not sufficient to show this, the main difference between active inference (viewed broadly so as to include maximum-entropy policy inference) and MOP seems to be the presence or absence of explicit model evidence constraints. 

The core concept in MOP is that maximizing path occupancy is an ``intrinsic'' value, from which reward is derivative. The core claim of the FEP and active inference (which we have seen to entail empowerment) is that maximizing \emph{model evidence} is an ``intrinsic'' value, and that rewards as well as information-seeking behavior derive from this imperative. At first glance, these frameworks may appear difficult to reconcile, since the former maximizes surprisal while the latter minimizes it (at least with respect to sensory observations).

One of the central claims of \citep{RamirezRuiz2022ComplexBF} is that intelligent, goal-directed action emerges naturally from the MOP objective, in the presence of absorbing states together with the means to (foreseeably) avoid them given certain courses of action. It may be wondered whether pure MOP agents would be as successful in less predictable environments in which risk-aversion may be more important, but independently of this, there are deep reasons to suppose that MOP agents would not produce richly intelligent behavior without an implicit model-evidence constraint. 

Occupancy-maximizing agents seek control only in order to remain alive, a goal that is argued to flow elegantly from the desire to maximize entropy in the distant future. However, this argument assumes that being dead corresponds to an ``absorbing'' state, which in the experiments is modeled as entailing zero entropy for the rest of time. In a more physically realistic model, dying would correspond to a breakdown of the agent-environment boundary, and so to a much \emph{higher}-entropy state (with the dissolution of individual agents corresponding to an unconstrained maximum-entropy state, or in physical terms, thermal equilibrium). Relatedly, the ``survival instinct'' is encoded in active inference agents in the fact that departures from homeostatic set points (defined by the generative model or ``preference distribution'') score high in free energy and so are aversive.

Thus, identifying a lack of action availability with a low-entropy state is plausible only in toy scenarios in which the entropy increase induced within the overall system by the dissolution of the agent is ignored. Death ought to be \emph{attractive} to MOP agents unless they possess an \emph{a priori} distinction between agent and environment, i.e. a ``sense of self''. The upshot is that the implicit constraint enabling the emergence of goal-directed behavior in such agents is, in the general case, not simply long-term entropy maximization but also the existence of an agent with a repertoire of actions, encoded in the very partitioning of the space into action and state variables. Effectively, this amounts to a version of the ``controllability'' constraints that appear explicitly in active inference and empowerment, as the agent must exert control sufficient to enable homeostasis (i.e. the maintenance of internal states against dissipative forces).\footnote{We may note that the ability to predict the entropy of future states so as to compute state value---however this is implemented---also corresponds to some local disequilibrium and thus imposes a \emph{de facto} constraint on entropy (here, we are explicitly considering the entropy associated with the internal states of the agent).}

\subsection{Model evidence and the will to live}

Despite the arguments just given, the inversion of traditional assumptions about the relationship between exploration and reward highlighted by the MOP is appealing, as entropy maximization (albeit under constraints) appears to be an essential feature of intelligence and life \citep{10.1063/1.4818538, PhysRevLett.110.168702, sakthivadivel2022geometryanalysisbayesianmechanics, dacosta2024probabilisticprinciplesbiophysicsneuroscience}, more constant across distinct forms of life than any particular reward-seeking behavior. The idea that future path occupancy, as measured by entropy \citep{RamirezRuiz2022ComplexBF}, is tantamount to remaining alive is one way of understanding the place of entropy maximization at the heart of accounts of intrinsic motivation.

We have seen however that in order to reproduce the goal-oriented behavior characteristic of complex biological intelligence, it is necessary to maximize entropy under the constraint that the agent's existence, operationalized as a conditional independence between internal and external states \citep{Friston2019AFE} (which appears in simple models as an action-state partition), is maintained. Taking a page from Schopenhauer \citep{Schopenhauer1958-SCHTWA-20}, intrinsic motivation may then be cast as simply the ``will to live'', i.e. to persist as a living (moving, changing) thing, a basal impulse that takes different particular forms depending on local constraints (generative models). These constraints shape the primary motivational force of entropy production, such that conditional independence structures are maintained. 

In simpler models of intelligence, the relevant partitioning of the entire (agent-environment) system is assumed to be fixed, but in more sophisticated treatments such as multi-scale or scale-free active inference \citep{10.1007/978-3-030-00075-2_7, friston2024pixelsplanningscalefreeactive}, model structure itself may evolve, typically at slower timescales. We may then view the life of an agent at any given instant as seeking not only observational evidence for the currently parameterized model, but also evidence for the parameters themselves, as well as for hyperparameters (or priors over parameters, including structural priors). This structural evolution can be understood in terms of Bayesian model selection \citep{FRISTON2024108891}. 

From this perspective, there is no deep contradiction between scale-free self-evidencing (i.e. the seeking of model evidence) \citep{Hohwy2014-HOHTSB-2} and maximum occupancy. Once constraints (parameters and model structure) are themselves treated as random variables, the process of self-evidencing is seen to be data- or observation-driven through and through, and it appears to be a property of our universe (insofar as it is accurately modeled as a closed system) that the entropy of data-generating processes as a whole can only increase. From this perspective, maximum-entropy inference is a ubiquitous self-fulfilling prophecy in virtue of which the universe evolves toward thermal equilibrium.\footnote{The distinction between thermodynamic and merely information-theoretic or variational free energy \citep{Kiefer2020-KIEPIA, e26080622} needn't concern us here, as the entropy of observations is sufficient to drive this process.} Thus all agents indeed maximize occupancy on the longest timescale, though in a rather selfless way, i.e. they gather evidence for a maximum-entropy model of the universe at large, in which boundaries between agents (Markov blankets) and their corresponding energetic constraints have disappeared.

The idea that entropy is maximized ``for its own sake'' does not, of course, preclude interpretations of this phenomenon in terms of epistemic value \citep{Friston2015ActiveIA}, curiosity \citep{Schmidhuber2010FormalTO}, and so on, in various contexts. What the preceding discussion does suggest is that exploratory behavior is by no means ``merely'' an evolved mechanism for securing outcomes high in utility, but is at least as fundamental an aspect of agency as the latter tendency, with the two plausibly participating in a dance of circular causality. The presence of both goal-seeking and information-seeking drives in the expected free energy functional, regardless of the particular generative model, points to this same conclusion \citep{Smith2022-SMIAIM-4}.

\section{Conclusion}

Seeking common themes across contemporary accounts of intrinsic motivation has surfaced the inevitability of constrained entropy maximization as a core principle describing motivation in biological systems. This insight is hardly novel at a fundamental level, as entropy maximization has long been recognized as a crucial principle both in physics generally \citep{PhysRev.106.620} and for the physics of life and intelligence specifically \citep{PhysRevLett.110.168702, Friston2019AFE, Ueltzhffer2020OnTT}, and has played an explicit role in several accounts of intrinsic motivation \citep{Schmidhuber2010FormalTO, PRXLife.2.033009}. Here, the goal has been primarily to explore in detail how three accounts of intrinsic motivation that have previously been juxtaposed in the literature \citep{moreno-bote2023empowerment} may nonetheless be understood as variants of this general perspective.

\acknowledgments{The author would like to thank in particular Karl Friston, Jacqueline Hynes, and Dalton Sakthivadivel for conversations directly relevant to this work, as well as Mahault Albarracin, Riddhi J. Pitliya, Maxwell Ramstead, Tim Verbelen, and Ran Wei for inspiring discussions.}

\reftitle{References}


\bibliography{main.bib}

\end{document}